\documentstyle[12pt]{article}
\begin{document}
\begin{center}
{\large\bf Absorption of TeV Photons and $\kappa$-deformed Poincar\'{e} 
algebra\footnote{Supported in part by KBN grant, 5PO3B05620}}
\end{center}
\vskip 1.5 cm
\begin{center}
{\bf Giovanni AMELINO-CAMELIA}$^a$, {\bf Jerzy LUKIERSKI}$^b$ and 
{\bf Anatol NOWICKI}$^c$\\
\end{center}
\noindent
$~~~~~~~~~~$
{\it $^a$Dipart.~Fisica,
Univ.~Roma ``La Sapienza'', and INFN Sez.~Roma1\\
\noindent
$~~~~~~~~~~$$~$
P.le Moro 2, 00185 Roma, Italy}\\
\noindent
$~~~~~~~~~~$
{\it $^b$Institute for Theoretical Physics, 
University of Wroc{\l}aw,\\
\noindent
$~~~~~~~~~~$$~$
pl. Maxa Borna 9, 50--205 Wroc{\l}aw, Poland}\\
\noindent
$~~~~~~~~~~$
{\it $^c$Institute of Physics, Pedagogical University\\
\noindent
$~~~~~~~~~~$$~$
pl. S\l owia\'{n}ski 6,
65-069 Zielona G\'{o}ra, Poland}

\vspace{1cm}
\begin{center}
{\bf ABSTRACT}
\end{center}

{\leftskip=0.6in \rightskip=0.6in 
  
We consider the process of collision between a hard photon and a
soft photon producing an electron-positron pair,
under the assumption that the kinematics be described 
according to the $\kappa$-deformation of the D=4 Poincar\'{e} algebra.
We emphasize the relevance of this analysis for the understanding
of the puzzling observations of multi-TeV photons from Markarian 501.
We find a significant effect of the $\kappa$-deformation
for processes above threshold, while, in agreement with a previous study,
we find that there is no leading-order deformation of the threshold condition.Š
}


\newpage

Recently, there has been increasing interest in tests of
phenomenological models describing effects with magnitude set
by the Planck length
(see, {\it e.g.}, Refs.~\cite{grbgac,gampul,gacgwi,daPLA}).
In particular, following the proposal put forward in
Ref.~\cite{grbgac},  Kluzniak~\cite{kluz} (also see the later Ref.~\cite{ita})
suggested that the universe might be more transparent to multi-TeV photons
than expected by conventional relativistic astrophysics. 
In fact, the phenomenological model of quantum-gravity ``space-time foam" 
considered in Ref.~\cite{grbgac}
has significant implications~\cite{gactp} for the mechanism of 
hard-photon disappearence in the far infrared  background  (FIRB)
through electron-positron pair production.
It was pointed out already in Ref.~\cite{stan} that
if the spectrum of photons observed from distant
astrophysical sources goes significantly beyond 10 TeV
we would have to ``revise our concepts about the propagation of TeV
gamma-rays in intergalactic space". 
Recent FIRB data from DIRBE~\cite{Wright00,Finkbeiner,Hauser98} and
from ISOCOM~\cite{Biviano99} have been used to render more
stringent this 10-TeV limit.
As observed by Protheroe and Meyer~\cite{aus},
this result raises a puzzle in light of the fact that
HEGRA has detected~\cite{Aharonian99} photons with a spectrum 
ranging up to 24 TeV from Markarian 501 
(a BL Lac object at a distance of $\sim 150$Mpc).

As shown in Ref.~\cite{gactp} the space-time foam model of Ref.~\cite{grbgac}
provides a fully consistent explanation of this paradox.
We are here interested in an appealing alternative solution of the paradox,
based on the role of the Planck length in quantum deformations of the 
Poincar\'{e} algebra, rather than in the structure of space-time foam.
In particular, we focus on one of 
the $\kappa$-Poincar\'{e}~\cite{kpoinold,nowi10,kpoinannal}
Hopf algebras\footnote{We recall that~\cite{drinf,woro,majbook}
noncocommutative Hopf algebras characterize quantum deformations 
of Lie groups and Lie algebras.}, 
in which the deformation parameter, $\kappa$,
can be naturally (although not necessarily) associated
with the Planck length: $\kappa \sim 1/L_p$.
A preliminary analysis of the paradox in terms of $\kappa$-Poincar\'{e}
was already reported in Ref.~\cite{gactp}. The simplest interpretation
of the paradox is in terms of a ``threshold anomaly" (a deformed
threshold condition for the relevant process) and it might have been
desireable to find a significant threshold anomaly following 
from $\kappa$-Poincar\'{e};
however, it was shown in Ref.~\cite{gactp,gacdsr} that the $\kappa$-Poincar\'{e}
threshold anomaly is negligibly small. Still, one can explore other
aspects of $\kappa$-Poincar\'{e} in relation with the Markarian-501 paradox.
In fact, the evaluation of the optical depth is not only sensitive
to the threshold condition: it is also sensitive~\cite{gactp,gacdsr} 
to the nature of processes somewhat above threshold.
It is in this respect that the analysis here reported may turn out
to have significant implications for the Markarian-501 paradox:
in fact, we show that the kinematics of electron-positron pair
production above threshold is significantly affected
by the $\kappa$ deformation.

Because of the significance of the experimental context 
we make our point within the simplest
(least technical) supporting analysis: head-on collision with equipartition
of outgoing energy, focusing mostly on the leading
order in $1/\kappa$, which is the natural expansion parameter
for $\kappa$-Poincar\'{e} analyses.
We also focus on one particular form of $\kappa$-deformed Poincar\'{e} 
algebra, written in the 
so-called ``bicrossproduct basis"~\cite{nowi10,kpoinannal}.
Moreover, in our analysis we adopt 
the conventional quantum-group framework\footnote{Alternative
proposals for the role of the coproduct in the kinematics of collision 
processes have been considered in the context of the theory
being developed in Refs.~\cite{gacdsr,jurekdsr,gacmarz}.} for the role
of the $\kappa$-Poincar\'{e} coproduct~\cite{kpoinold,nowi10,kpoinannal}
in the kinematics of collision processes.
The only other $\kappa$-Poincar\'{e} structure
that is relevant for the determination of the $\kappa$-deformed
kinematic rules for collision processes is the $\kappa$-deformed
mass-shell condition~\cite{kpoinold,nowi10,kpoinannal}
(dispersion relation).

Let us therefore start by noting here the known formulas
for the mass shell and the coproduct.
In bicrossproduct basis~\cite{nowi10,kpoinannal}
the $\kappa$-Poincar\'{e} mass-shell condition is (we set $c=1$ throughout)
\begin{equation}  \label{now3}
\left(2 \kappa \, \sinh {\frac{p_{0} }{2\kappa}}
\right)^{2}- \overrightarrow{p}^{2} e^{{\frac{p_{0}}{\kappa}}} 
= \left(2 \kappa \, \sinh {\frac{m}{2\kappa}}
\right)^{2} 
\end{equation}
and a consistent rule for composition of energy-momentum (coproduct
rule) is~\cite{nowi10,kpoinannal}
\begin{eqnarray}
\qquad\overrightarrow{p}^{(1+2)} = \overrightarrow{p}^{(1)} +
 e^{- {\frac{p^{(1)}_{0}}{\kappa}}} \, \overrightarrow{p}^{(2)}
\label{szea}
\end{eqnarray}
\begin{eqnarray}
\qquad {p}_0^{(1+2)} = {p}_0^{(1)} + {p}_0^{(2)} ~. 
\label{szeener}
\end{eqnarray}

Incidentally, we remind the reader that 
the non--Abelian (noncocommutative) addition law
of three-momenta corresponds to the deformation of the dual
space--time picture. The noncommutative dual Minkowski space-time
coordinates $\widehat{x}_{\mu} = ( \widehat{x}_{i}, \widehat{x}_{0})$ 
(see~\cite{nowi10,kpoinannal,zak}) satisfy  the relations
\begin{equation}
\left[ \widehat{x}_{0},\widehat{x}_{i}\right] =
{\frac{i}{\kappa}} \widehat{x}_{i} \qquad \left[
\widehat{x}_{i}, \widehat{x}_{j}\right] = 0 ~.
\end{equation}

We are now ready for the analysis of the $\kappa$-deformed kinematics of 
$\gamma + \gamma \to e^{+} + e^{-}$. The case of interest for the
Markarian-501 paradox involves the collision of a hard (energy $E$,
momentum $\vec{{P}}$) photon
and a soft (energy $\epsilon \ll E$, momentum $\vec{{p}}$) photon.
We denote with $\vec{{p}}_+, E_+$ and $\vec{{p}}_-, E_-$
the energy-momentum of the outgoing electron-positron pair.
As announced we focus on head-on collisions 
($\vec{{P}} {\cdot} \vec{{p}} = P p$) and we adopt 
the conventional framework~\cite{kpoinold,nowi10,kpoinannal} 
for the role of the $\kappa$-Poincar\'{e} coproduct
in the kinematics of collision processes, which amounts to the
conservation of the coproduct sum of energy-momentum:
\begin{eqnarray}
\qquad \overrightarrow{p}^{(1+2)}_{in} = \overrightarrow{p}^{(1+2)}_{out}
\label{consp}
\end{eqnarray}
\begin{eqnarray}
\qquad {p}_{0,in}^{(1+2)} = {p}_{0,out}^{(1+2)} ~.
\label{conse}
\end{eqnarray}

It is also convenient to introduce the angle $\theta$ between the two outgoing
momenta: $\vec{p}_+  {\cdot} \vec{p}_- = {p}_+ {p}_- \cos\theta$. 
In fact, one can use (\ref{consp}) to obtain a relation
between the square-moduli, 
$\overrightarrow{p}^{(1+2)}_{in} {\cdot} \overrightarrow{p}^{(1+2)}_{in} = 
\overrightarrow{p}^{(1+2)}_{out} {\cdot} \overrightarrow{p}^{(1+2)}_{out}$,
which takes the form
\begin{eqnarray}
P^2- 2 P p \simeq p_+^2 + p_-^2 + 2 p_+ p_- \cos\theta 
- {2 \over \kappa} E_+ p_-^2 - {2 \over \kappa} E_+ p_+ p_- \cos\theta ~,
\label{squaremodul}
\end{eqnarray}
where we used the rule (\ref{szea})
with the identifications
$\overrightarrow{p}^{(1)}_{in} =  \vec{P}$, 
$\overrightarrow{p}^{(2)}_{in} =  \vec{p}$, 
$\overrightarrow{p}^{(1)}_{out} =  \vec{p}_+$, 
$\overrightarrow{p}^{(2)}_{out} =  \vec{p}_-$;
moreover, as announced, we are including only the 
leading-order $\kappa$-dependent corrections, neglecting terms
which, in addition to the $1/\kappa$ suppression,
are also suppressed by the smallness of $\epsilon$.
Also a term of order $p^2$ has been dropped on the left-hand side
(in the $\gamma + \gamma \to e^{+} + e^{-}$ processes relevant
for the Markarian-501 paradox the energy/momentum scales
are such that $p^2 \ll p_{\pm}^3/\kappa$).

Analogously, it is convenient to square (\ref{conse}) obtaining
\begin{eqnarray}
E^2 + 2 E \epsilon \simeq E_+^2 + E_-^2 + 2 E_+ E_-
~.
\label{squarener}
\end{eqnarray}

Up to this point we have only used (coproduct-modified)
energy-momentum conservation. We must now enforce 
the $\kappa$-Poincar\'{e} mass-shell condition. In the case being
studied from (\ref{now3}) it follows that
\begin{eqnarray}
E \simeq P + {1 \over 2 \kappa} P^2 ~,~~~ 
\epsilon \simeq p ~,~~~ 
E_{\pm} \simeq p_{\pm} + {m_e^2 \over 2 p_{\pm}}+ {1 \over 2 \kappa} p_{\pm}^2 ~,
\label{disprels}
\end{eqnarray}
where, again, 
we only included the 
leading-order $\kappa$-dependent corrections
and in addition we also took into account (even in the $\kappa$-independent
terms) the smallness of $m_e$ with respect to $p_+$ and $p_-$
(certainly well justified for TeV electrons).

Eq.~(\ref{disprels}) allows to 
rewrite (\ref{squaremodul}) and (\ref{squarener}) as
\begin{eqnarray}
P^2- 2 P \epsilon \simeq p_+^2 + p_-^2 + 2 p_+ p_- \cos\theta 
- {2 \over \kappa} p_+ p_-^2 - {2 \over \kappa} p_+^2 p_- \cos\theta ~,
\label{squaremodulbis}
\end{eqnarray}
\begin{eqnarray}
P^2 + 2 P \epsilon + {1 \over \kappa} P^3 \simeq 
\left( p_+ + p_- + {m_e^2 \over 2 p_+} + {m_e^2 \over 2 p_-} 
+ {1 \over 2 \kappa} p_+^2
+ {1 \over 2 \kappa} p_-^2 \right)^2 ~,
~.
\label{squarenerbis}
\end{eqnarray}
These equations (\ref{squaremodulbis}) and (\ref{squarenerbis}) 
establish, for given value of $\epsilon$, two relations
between $P,p_+,p_-,\theta$.
We observe that Eqs.~(\ref{squaremodulbis}) and (\ref{squarenerbis})
show that, 
while the leading order $\kappa$-deformation
of the threshold condition vanishes~\cite{gactp}, 
the kinematic conditions for generic processes above threshold 
are affected by the $\kappa$-deformation in leading order.
In order to show this feature more explicitly, as announced, we focus
on the particular case in which the outgoing energy is equipartitioned
between the electron and the positron (also because, in particular, 
equipartition applies to threshold electron-positron pair production).
We can therefore, for the remainder of this paper,
adopt the notations $E' \equiv E_+ = E_-$ and $p' \equiv p_+ = p_-$.
This allows us to rewrite (\ref{squaremodulbis}) and (\ref{squarenerbis}),
for the case of equipartition of outgoing energy, as
\begin{eqnarray}
P^2- 2 P \epsilon \simeq 2 p'^2 (1+ \cos\theta) 
- {2 \over \kappa} p'^3 (1 + \cos\theta) ~,
\label{squaremodultris}
\end{eqnarray}
\begin{eqnarray}
P^2 + 2 P \epsilon + {1 \over \kappa} P^3 \simeq 
4 p'^2 + 4 m_e^2 + {4 \over \kappa} p'^3 ~,
~.
\label{squarenertris}
\end{eqnarray}

The physical content of this result becomes more visible if we 
combine Eqs.~(\ref{squaremodultris}) and (\ref{squarenertris})
to obtain the following equivalent (again, to leading order) 
kinematical conditions
\begin{eqnarray}
p'^2 \simeq {P^2 \over 3 + \cos \theta} - {2 m_e^2 \over 3 + \cos \theta}
+ {1 - 2 (1 - \cos \theta) (3 
+ \cos \theta)^{-3/2} \over 6 + 2 \cos \theta} {P^3 \over \kappa} 
\label{pokerpp}
\end{eqnarray}
\begin{eqnarray}
P &\! \simeq \!& \left( 1 -  {1- \cos \theta \over 3 + \cos \theta} \right) 
{m_e^2 \over \epsilon} 
+ {2 (1 - \cos \theta) \over 3 + \cos \theta} {P^2 \over 4 \epsilon}
+ \nonumber\\
&~& + \left( {1- \cos \theta \over 3 + \cos \theta}
- 1 + {4 (3 + \cos \theta)^2 -4 (1 - \cos \theta)^2
\over 2 (3+ \cos \theta)^{5/2}} \right) {P^3 \over 4 \epsilon \kappa}
~.
\label{pokerthresh}
\end{eqnarray}
Eq.~(\ref{pokerthresh}) is our key result. It is a self-consistent equation
that establishes the value of the momentum $P$ of the hard photon
required for a head-on collision with a soft-photon of given energy $\epsilon$
to produce an electron-positron pair with equipartition of energy
and with a given value of the angle $\theta$ 
between $\vec{{p}}_+$ and $\vec{{p}}_-$. The following observations 
are in order:

(i) In the classical-space-time limit ($\kappa \rightarrow \infty$)
our formula (\ref{pokerthresh}) of course reproduces
the corresponding classical special-relativistic kinematical conditions
for pair production with equipartition of outgoing energy.

(ii) For pair production at threshold ($\theta = 0$)
our result predicts an exact
cancellation among the leading-order $\kappa$-dependent terms,
confirming the earlier result reported in Refs.~\cite{gactp,gacdsr}.

(iii) For pair production above threshold our result predicts
a non-vanishing leading-order $\kappa$-dependent correction and this 
correction is such that, for given soft-photon energy $\epsilon$,
the process would require hard-photon momentum/energy that is 
higher\footnote{It is known~\cite{kpoinannal} 
that there are two versions 
of the $\kappa$-deformed Poincar\'{e} algebra
in the bicrossproduct basis, differing by the sign in front of 
the $\kappa$ parameter. 
The results here obtained also apply to the other version 
of the bicrossproduct basis, upon replacing 
consistently $\kappa \rightarrow - \kappa$,
in which case, of course, the effect we found goes in the opposite
direction: for given soft-photon energy the process would require
hard-photon momentum/energy that is smaller than the corresponding
prediction of classical special-relativistic kinematics.}
than the corresponding prediction of classical special-relativistic kinematics.
Even for $\kappa$ large enough to
satisfy $\kappa \sim 1/L_p$, the contribution to $P$ coming from the term
of order $P^3/(\epsilon \kappa)$ can be quite significant ($P/\kappa$ is
very small, but $P/\epsilon$ is very large~\cite{gactp}).

(iv) We also emphasize that, while at the coproduct
level the identifications $\overrightarrow{p}^{(1)}_{out} =  \vec{p}_+$, 
$\overrightarrow{p}^{(2)}_{out} =  \vec{p}_-$ are not equivalent
to the identifications $\overrightarrow{p}^{(1)}_{out} =  \vec{p}_-$, 
$\overrightarrow{p}^{(2)}_{out} =  \vec{p}_+$,
the final result of our analysis
does not depend on this choice.

We leave for future studies (readers from the astrophysics community
might be best equipped for this delicate phenomenological analysis)
the task of establishing whether
the correction we found for pair-production above threshold is sufficient
to explain the Markarian-501 paradox.
The ingredients for obtaining such an explanation are clearly present
in our result: in fact, as mentioned,
the evaluation of the optical depth is sensitive~\cite{gactp,gacdsr} 
to the nature of processes somewhat above threshold (actually, even
the peak of the pair-production cross section is somewhat above 
threshold~\cite{gactp}), and our result suggests that the phase space
available for pair-production by a hard photon with given energy $E$
might be reduced in $\kappa$-Poincar\'{e}, thereby allowing for a
modification of the optical depth result.

If indeed future studies will confirm that our result provides a solution
of the Markarian-501 paradox it should also be possible to distinguish
between this model and other models being considered for a solution
of the paradox. The mentioned space-time-foam model of Ref.~\cite{grbgac}
would interpret the paradox as a ``threshold anomaly" while in the model
here adopted 
the deformation of the threshold condition is negligible (but the deformation
of processes above threshold is significant). These two alternative
pictures should lead to different observable consequences, possibly
verifiable in future more refined experiments. 

Even easier is the
discrimination between the model here adopted and the Coleman-Glashow
Lorentz-invariance-violation model~\cite{colgla} 
which is also known~\cite{gactp,glaste}
to provide a solution to the Markarian-501 paradox.
In fact, the Coleman-Glashow model predicts deviations from
classical special-relativistic kinematics even in the low-energy
regime, whereas in the model here adopted the deformation
is completely negligible in low-energy phenomena 
(if indeed $\kappa \sim 1/L_p$). Sensitive tests of special-relativistic
kinematics for low-energy processes could therefore discriminate 
between the two models.

Whether or not the model here adopted does lead to a solution of 
the Markarian-501 paradox (and whether or not it proves to be
a better solution than its alternatives),
our result would remain useful as a characterization
of $\kappa$-Poincar\'{e} Hopf algebras. In almost a decade~\cite{kpoinold}
of research on this subject a large number of results have been
obtained, establishing the mathematical properties of the formalism,
but only very few characteristic predictions have been identified.
The hope of finding an explanation to the Markarian-501 paradox has
reenergized research in this direction. We established in this note
that a characteristic feature of $\kappa$-deformed kinematics
is the presence of leading-order corrections to the kinematic
rules for processes above threshold, while, as already 
shown in Ref.~\cite{gactp}, there is no
leading-order corretion to the threshold
condition.

\section*{Acknowledgements}
The authors would like to thank W.~Klu\'{z}niak and
J.~Kowalski-Glikman for valuable comments.


\end{document}